\documentclass[twocolumn,nopacs,floatfix,amsmath,nofootinbib,amssymb,preprintnumbers,floatfix]{revtex4}
\usepackage{amsmath}
\usepackage{longtable,lscape,booktabs}
\usepackage{txfonts}
\usepackage{overpic}
\usepackage{amssymb}
\usepackage{color}
\usepackage{longtable,lscape,bm}
\usepackage{multirow}
\usepackage{graphicx,color,dcolumn,bm}
\usepackage[colorlinks,
            citecolor=blue,
            anchorcolor=red,
            menucolor=red,
            linkcolor=red,
            filecolor=red,
            runcolor=red,
            urlcolor=blue,
            frenchlinks=red]{hyperref}

\renewcommand{\arraystretch}{1.50}
\allowdisplaybreaks

\begin{document}

\title{Doubly charmed molecular pentaquarks}

\author{Rui Chen$^{1}$}\email{chen$_$rui@pku.edu.cn}
\author{Ning Li$^{2}$} \email{lining59@mail.sysu.edu.cn}
\author{Zhi-Feng Sun$^{3, 4, 5}$} \email{sunzf@lzu.edu.cn}
\author{Xiang Liu$^{3,4,5}$} \email{xiangliu@lzu.edu.cn}
\author{Shi-Lin Zhu$^{1}$} \email{zhusl@pku.edu.cn}

\affiliation{ $^1$Center of High Energy Physics, Peking University,
Beijing
100871, China\\
$^2$School of Physics, Sun Yat-Sen University, Guangzhou 510275, China\\
$^3$School of Physical Science and Technology, Lanzhou University, Lanzhou 730000, China\\
$^4$Research Center for Hadron and CSR Physics, Lanzhou University and Institute of Modern Physics of CAS, Lanzhou 730000, China\\
$^4$Lanzhou Center for Theoretical Physics, Key Laboratory of
Theoretical Physics of Gansu Province, and Frontiers Science Center
for Rare Isotopes, Lanzhou University, Lanzhou 730000, China}

\begin{abstract}

We perform a systematic exploration of the possible doubly charmed
molecular pentaquarks composed of $\Sigma_c^{(*)}D^{(*)}$ with the
one-boson-exchange potential model. After taking into account the
$S-D$ wave mixing and the coupled channel effects, we predict
several possible doubly charmed molecular pentaquarks, which include
the $\Sigma_cD$ with $I(J^P) = 1/2(1/2^-)$, $\Sigma_c^*D$ with
$1/2(3/2^-)$, and $\Sigma_cD^*$ with $1/2(1/2^-)$, $1/2(3/2^-)$. The
$\Sigma_cD$ state with $3/2(1/2^-)$ and $\Sigma_cD^*$ state with
$3/2(1/2^-)$ may also be suggested as candidates of doubly charmed
molecular pentaquarks. The $\Sigma_cD$ and $\Sigma_c^*D$ states can
be searched for by analyzing the $\Lambda_cD\pi$ invariant mass
spectrum of the bottom baryon and $B$ meson decays. The
$\Sigma_cD^*$ states can be searched for in the invariant mass
spectrum of $\Lambda_cD^*\pi$, $\Lambda_cD\pi\pi$ and
$\Lambda_cD\pi\gamma$. Since the width of $\Sigma_c^*$ is much
larger than that of $D^*$, $\Sigma_c^*D\rightarrow \Lambda_cD\pi$
will be the dominant decay mode. We sincerely hope these candidates
for the doubly charmed molecular pentaqurks will be searched by LHCb
or BelleII collaboration in the near future.
\end{abstract}

\maketitle

\section{introduction}

In the past decades, the observations of $X/Y/Z/P_c$ states have
stimulated theorist's extensive interest in exploring the properties
of exotic states. Among these possible configurations of exotic
state, the hadronic molecular state is composed of the color-singlet
hadrons, which is different from other exotic state configurations,
like the hybrid, glueball, multiquarks. Since many observed
$X/Y/Z/P_c$ states are near the threshold of one hadron pair, the
molecular assignments have received extensive attentions
\cite{Chen:2016qju,Liu:2013waa,Guo:2017jvc,Hosaka:2016pey}. The
study of hadronic molecular state is an active and important
research field in hadron physics.

Among these extensive studies of charmoniumlike $XYZ$ states and
$P_c$ state, we may find shaped integrated and clear venation. The
charmoniumlike $XYZ$ states inspired the discussion of the
interaction between charmed meson and anti-charmed meson
\cite{Chen:2016qju,Liu:2013waa,Guo:2017jvc}. Here, a typical example
is that the $D\bar{D}^*$ molecular explanation of the $X(3872)$ was
proposed
\cite{Swanson:2003tb,Liu:2008fh,Lee:2009hy,Liu:2009qhy,Li:2012cs,Suzuki:2005ha,Thomas:2008ja},
which has been viewed as a starting point of exploring the hadronic
molecular tetraquark state since 2003. From these studies, the
applicability and reliability of the involved phenomenological
models like the one-boson exchanged model adopted in this work were
also tested. The authors in Refs.
\cite{Yang:2011wz,Wu:2010jy,Wang:2011rga,Karliner:2015ina,Li:2014gra,Wu:2010vk}
benefitted from their experience with the hadronic molecular
tetraquark states and further investigated the interaction between
the charmed meson and anti-charmed baryon and predicted the
hidden-charm pentaquarks. In 2015, LHCb reported the observation of
several $P_c$ states \cite{Aaij:2015tga}, which are consistent with
the prediction of the hidden-charm pentaquarks. In 2019, with more
precise data, LHCb again analyzed the same process and found the
characteristic mass spectrum of the $P_c$ states
\cite{Aaij:2019vzc}, which provides direct evidence of the existence
of the hidden-charm molecular pentaquarks
\cite{Chen:2019asm,Liu:2019tjn,He:2019ify,Xiao:2019aya,Meng:2019ilv,Yamaguchi:2019seo,Valderrama:2019chc,
Chen:2019bip,Burns:2019iih,Du:2019pij,Wang:2019ato}.

Very recently, the LHCb Collaboration analyzed the $D^0D^0\pi^+$
mass spectrum using the full Run1 plus Run2 data corresponding to 9
fb$^{-1}$, and observed a very narrow doubly charmed tetraquark
$T_{cc}^+$ as its minimal valence quark component is
$cc\bar{u}\bar{d}$~\cite{Tcc:talk}. Its mass relative to the
$D^0D^{*+}$ mass threshold and decay width are
\begin{eqnarray}
\delta m &=& -273\pm 61\pm 5^{+11}_{-14}~\text{keV}/c^2,\nonumber\\
\Gamma &=& 410\pm 165\pm 43^{+18}_{-38}~\text{keV},\nonumber
\end{eqnarray}
respectively. The spin-parity is estimated as $1^+$. The mass and
spin-parity are well consistent with the prediction of the $DD^*$
doubly charmed molecular
state~\cite{Li:2012ss,Xu:2017tsr,Li:2021zbw}. In particular, after
considering the isospin breaking effects, the newly $T_{cc}$ state
can be assigned as the $S-$wave $D^0D^{*+}$ molecular state. We also
predict another doubly charmed $D^+D^{*0}$ resonance with its mass
around 3876 MeV \cite{Chen:2021vhg}.

\begin{figure}[!htbp]
\centering
\includegraphics[width=3.3in]{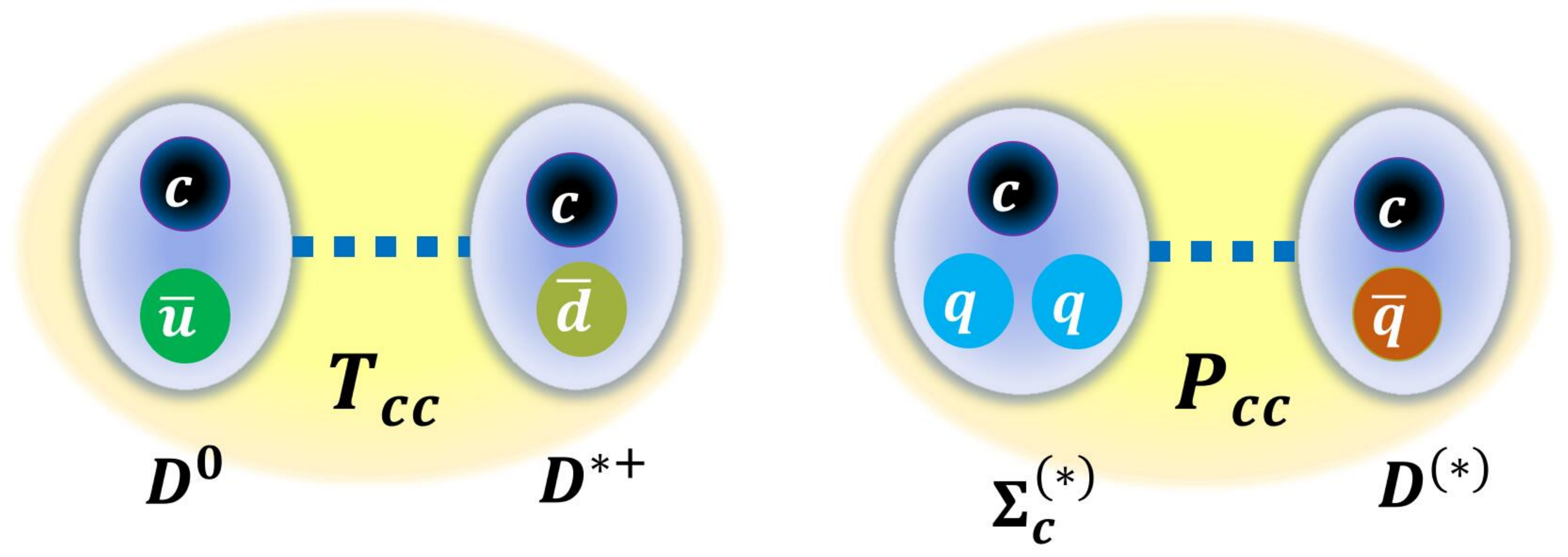}
\caption{A comparison between the $T_{cc}$ and $P_{cc}$ in the
doubly charmed molecular picture.} \label{tcc}
\end{figure}

The light quark pair within the $\Sigma_c^{(*)}$ baryon shares the
same color configuration with the light anti-quark within the $D^0$
meson. If the $T_{cc}$ is the doubly charmed molecule, there should
also exist the doubly charmed molecular pentaquarks composed of the
charmed baryon $\Sigma_c^{(*)}$ and charmed meson $D^{(*)}$ as shown
in Figure \ref{tcc}. The above argument is very similar to the
underlying reasoning of predicting the hidden-charm molecular
pentaquarks from the existence of the hidden-charm molecular states
\cite{Yang:2011wz}. The only difference is that now we have two
charm quarks instead of a $c\bar c$ pair.

Compared to the $DD^*$ system, there is one more light quark in the
$\Sigma_c^{(*)}D^{(*)}$ systems. If we neglect the very weak
interactions between the charmed quarks, the $\Sigma_c^{(*)}D^{(*)}$
interactions can be much stronger than the $DD^*$ interactions
\cite{Chen:2017vai}. In particular, we notice that the light quark
configuration for the $\Sigma_c^{(*)}D^{(*)}$ systems is
$(qq)-\bar{q}$, the $\Sigma_c^{(*)}D^{(*)}$ interactions can also be
much stronger (or less weaker) than the corresponding
$\Sigma_c^{(*)}\bar D^{(*)}$ interactions, while the $P_c$ states
observed by the LHCb Collaboration \cite{Aaij:2015tga,Aaij:2019vzc}
can be assigned as the $\Sigma_c^{(*)}\bar D^{(*)}$ hidden-charm
molecular pentaquarks
\cite{Chen:2019asm,Liu:2019tjn,He:2019ify,Xiao:2019aya,Meng:2019ilv,Yamaguchi:2019seo,Valderrama:2019chc,
Chen:2019bip,Burns:2019iih,Du:2019pij,Wang:2019ato}. Therefore, it
is well-motivated and very interesting to search for the possible
doubly charmed baryon-meson molecular states from the
$\Sigma_c^{(*)}D^{(*)}$ interactions.

In fact, there are several predictions on the doubly charmed
molecular pentaquarks \cite{Yang:2020twg,Chen:2021htr,Dong:2021bvy}.
For example, within the framework of chiral effective field theory
\cite{Chen:2021htr}, Chen {\it{et al}} performed a systematic study
on the interactions of the $\Sigma_c^{(*)}D^{(*)}$ interactions.
They found all the $S-$wave $\Sigma_c^{(*)}D^{(*)}$ systems with
isospin $I=1/2$ can be possible doubly charmed molecular
pentaquarks, and their binding energies are larger than the
corresponding $\Sigma_c^{(*)}\bar D^{(*)}$ bound states. In Ref.
\cite{Dong:2021bvy}, the authors obtained the similar conclusions in
the resonance saturation model.

In this work, we adopt the one-boson-exchange (OBE) model to derive
the effective potentials describing the $\Sigma_c^{(*)}D^{(*)}$
interactions, and consider the $S-D$ wave mixing effects and the
coupled channel effects. Our investigation will not only provide
valuable information to experimental search for the doubly charmed
molecular pentaquarks, but also give indirect test of the molecular
state picture for the $P_c$ and $T_{cc}^+$ states.

This paper is organized as follows. After the introduction, we
present the detailed deduction of the effective potentials for the
$\Sigma_c^{(*)}D^{(*)}$ systems in Sec. \ref{sec2}. In Sec.
\ref{sec3}, we present the corresponding numerical results by
solving the coupled channel Shr$\ddot{\text{o}}$dinger equation. The
paper ends with the summary in Sec. \ref{sec4}.

\section{Interactions}\label{sec2}

As a molecular state composed of two colorless hadrons, its wave
function is constructed by three parts, i.e., the flavor wave
function, the spin-orbit wave function, and the radial wave
function. Here, the flavor wave functions $|I, I_3\rangle$ for the
$\Sigma_c^{(*)}{D}^{(*)}$ systems are written as
\begin{eqnarray*}&&\begin{array}{c}
\left|\frac{1}{2},\frac{1}{2}\right\rangle =
     \sqrt{\frac{2}{3}}\left|\Sigma_c^{(*)++}{D}^{(*)0}\right\rangle
     +\frac{1}{\sqrt{3}}\left|\Sigma_c^{(*)+}{D}^{(*)+}\right\rangle,\\
\left|\frac{1}{2},-\frac{1}{2}\right\rangle =
     \frac{1}{\sqrt{3}}\left|\Sigma_c^{(*)+}{D}^{(*)0}\right\rangle
     +\sqrt{\frac{2}{3}}\left|\Sigma_c^{(*)0}{D}^{(*)+}\right\rangle,
     \end{array}\\
&&\begin{array}{l}
\left|\frac{3}{2},\frac{3}{2}\right\rangle = -\left|\Sigma_c^{(*)++}{D}^{(*)+}\right\rangle,\\
\left|\frac{3}{2},\frac{1}{2}\right\rangle =
     \frac{1}{\sqrt{3}}\left|\Sigma_c^{(*)++}{D}^{(*)0}\right\rangle-\sqrt{\frac{2}{3}}
     \left|\Sigma_c^{(*)+}{D}^{(*)+}\right\rangle,\\
\left|\frac{3}{2},-\frac{1}{2}\right\rangle =\sqrt{\frac{2}{3}}
    \left|\Sigma_c^{(*)+}{D}^{(*)0}\right\rangle- \frac{1}{\sqrt{3}}\left|\Sigma_c^{(*)0}{D}^{(*)+}\right\rangle,\\
\left|\frac{3}{2},-\frac{3}{2}\right\rangle =
     \left|\Sigma_c^{(*)0}{D}^{(*)0}\right\rangle.
     \end{array}
\end{eqnarray*}

When we consider the $S-D$ wave mixing and the coupled channel
effects, the spin-orbit wave functions $|{}^{2S+1}L_J\rangle$ for
the discussed $\Sigma_c^{(*)}D^{(*)}$ systems include
$\Sigma_cD|{}^4S_{1/2}\rangle$,
$\Sigma_c^*{D}|{}^4{D}_{1/2}\rangle$,
$\Sigma_c{D}^*|{}^2{S}_{1/2},{}^4{D}_{1/2}\rangle$,
$\Sigma_c^*{D}^*|{}^2{S}_{1/2},{}^4{D}_{1/2},{}^6{D}_{1/2}\rangle$
for the $J^P=1/2^-$ and
$\Sigma_c^*{D}|{}^4{S}_{3/2},{}^4{D}_{3/2}\rangle$,
$\Sigma_c{D}^*|{}^4{S}_{3/2}, {}^2{D}_{3/2}, {}^4{D}_{3/2}\rangle$,
$\Sigma_c^*{D}^*|{}^4{S}_{3/2}, {}^2{D}_{3/2}, {}^4{D}_{3/2},
{}^6{D}_{3/2}\rangle$ for the $J^P=3/2^-$, respectively. The general
expressions for the spin-orbit wave functions for the
$\Sigma_c^{(*)}{D}^{(*)}$ systems are constructed as
\begin{eqnarray}
\Sigma_cD:\,\, \left|{}^{2S+1}L_{J}\right\rangle &=&
\sum_{m_S,m_L}C^{J,M}_{\frac{1}{2}m_S,Lm_L}
          \chi_{\frac{1}{2}m_S}|Y_{L,m_L}\rangle,\nonumber\\
\Sigma_c^*{D}:\,\, \left|{}^{2S+1}L_{J}\right\rangle &=&
\sum_{m_S,m_L}C^{J,M}_{\frac{3}{2}m_S,Lm_L}
          \Phi_{\frac{3}{2}m_S}|Y_{L,m_L}\rangle,\nonumber\\
\Sigma_c{D}^*: \left|{}^{2S+1}L_{J}\right\rangle &=&
\sum_{m,m'}^{m_S,m_L}C^{S,m_S}_{\frac{1}{2}m,1m'}C^{J,M}_{Sm_S,Lm_L}
          \chi_{\frac{1}{2}m}\epsilon^{m'}|Y_{L,m_L}\rangle,\nonumber\\
\Sigma_c^*{D}^*: \left|{}^{2S+1}L_{J}\right\rangle &=&
\sum_{m,m'}^{m_S,m_L}C^{S,m_S}_{\frac{3}{2}m,1m'}C^{J,M}_{Sm_S,Lm_L}
          \Phi_{\frac{3}{2}m}\epsilon^{m'}|Y_{L,m_L}\rangle,\nonumber
\end{eqnarray}
where $C^{J,M}_{Sm_S,Lm_L}$, $C^{S,m_S}_{\frac{1}{2}m,1m'}$, and
$C^{S,m_S}_{\frac{3}{2}m,1m'}$ are the Clebsch-Gordan coefficients.
$\chi_{\frac{1}{2}m}$ and $Y_{L,m_L}$ stand for the spin wave
function and the spherical harmonics function, respectively. The
polarization vector $\epsilon$ for the $\bar{D}^*$ vector meson is
defined as
$\epsilon_{\pm}^{m}=\mp\frac{1}{\sqrt{2}}\left(\epsilon_x^{m}{\pm}i\epsilon_y^{m}\right)$
and $\epsilon_0^{m}=\epsilon_z^{m}$, which satisfy $\epsilon_{\pm1}=
\frac{1}{\sqrt{2}}\left(0,\pm1,i,0\right)$ and $\epsilon_{0}
=\left(0,0,0,-1\right)$. The polarization tensor
$\Phi_{\frac{3}{2}m}$ for $\Sigma_c^*$ is expressed as
$\Phi_{\frac{3}{2}m}=\sum_{m_1,m_2}\langle\frac{1}{2},m_1;1,m_2|\frac{3}{2},m\rangle
\chi_{\frac{1}{2},m_1}\epsilon^{m_2}$.

The OBE effective potentials for the $\Sigma_c^{(*)}{D}^{(*)}$
interactions can be related to the corresponding OBE effective
potentials for the $\Sigma_c^{(*)}\bar{D}^{(*)}$ systems by using
the $G$ parity rule, i.e.,
\begin{eqnarray}
V_{\Sigma_c^{(*)}{D}^{(*)}} &=&
(-1)^{G_{E}}V_{\Sigma_c^{(*)}\bar{D}^{(*)}}.\label{obe}
\end{eqnarray}
Here, $G_{E}$ stands for the $G$ parity for the exchanged mesons,
which include the scalar meson $\sigma$, pseudoscalar mesons
$\pi/\eta$, and vector mesons $\rho/\omega$.

In our previous work, we have deduced the concrete OBE effective
potentials for the $\Sigma_c^{(*)}\bar{D}^{(*)}$ systems by
employing the effective Lagrangian approach at the hadronic level.
The general procedures can be divided into three steps. We first
construct the effective Lagrangians relevant to the interactions
between the $S-$wave charmed baryons/mesons and light mesons
\cite{Yan:1992gz,Wise:1992hn,Burdman:1992gh,Casalbuoni:1996pg,Falk:1992cx},
and write down the corresponding scattering amplitudes
$\mathcal{M}_{\Sigma_c^{(*)}\bar{D}^{(*)}\to
\Sigma_c^{(*)}\bar{D}^{(*)}}$ for the $t$ channel
$\Sigma_c^{(*)}\bar{D}^{(*)}\to \Sigma_c^{(*)}\bar{D}^{(*)}$
processes by exchanging the light mesons. Secondly, we derive the
effective potentials in the momentum space by using the Breit
approximation, $\mathcal{V}_{E}^{h_1h_2\to h_3h_4}({\bf{q}}) =
-{\mathcal{M}_{h_1h_2\to h_3h_4}}/{\sqrt{\prod_i2M_i\prod_f2M_f}}$.
$M_i$ and $M_f$ are the masses of the initial states ($h_1$, $h_2$)
and final states ($h_3$, $h_4$), respectively. Finally, we perform
the Fourier transformation to obtain the effective potential in the
coordinate space $\mathcal{V}({\bf{r}})$, i.e.,
\begin{eqnarray}
\mathcal{V}_{E}^{h_1h_2\to h_3h_4}({\bf{r}}) =
          \int\frac{d^3{\bf{q}}}{(2\pi)^3}e^{i{\bf{q}}\cdot{\bf{r}}}
          \mathcal{V}_{E}^{h_1h_2\to h_3h_4}({\bf{q}})\mathcal{F}^2(q^2,m_E^2).\nonumber
\end{eqnarray}
In the above formula, we introduce a monopole type form factor
$\mathcal{F}(q^2,m_E^2)= (\Lambda^2-m_E^2)/(\Lambda^2-q^2)$ at every
interactive vertex to compensate the off-shell effects of the
exchanged boson. $\Lambda$, $m_E$, and $q$ are the cutoff, the mass
and four-momentum of the exchanged meson, respectively.

According to the relations in Eq. (\ref{obe}) and the expressions in
Ref. \cite{Chen:2019asm}, we finally obtain the OBE effective
potentials for the
$\Sigma_c^{(*)}{D}^{(*)}\to\Sigma_c^{(*)}{D}^{(*)}$ processes as
summarized in Table \ref{potentials}. And we define several useful
functions, i.e.,
\begin{eqnarray}
Y(\Lambda,m,{r}) &=&\frac{1}{4\pi r}(e^{-mr}-e^{-\Lambda r})-\frac{\Lambda^2-m^2}{8\pi \Lambda}e^{-\Lambda r},\\
\mathcal{Y}^{ij}_{\Lambda m_a}&=&\mathcal{D}_{ij}Y(\Lambda,m_\sigma,r),\\
\mathcal{Z}^{ij}_{\Lambda m_a}&=&\left(\mathcal{E}_{ij}\nabla^2+\mathcal{F}_{ij}r\frac{\partial}{\partial r}\frac{1}{r}\frac{\partial}{\partial r}\right)Y(\Lambda,m_a,r),\\
\mathcal{Z}^{\prime ij}_{\Lambda
m_a}&=&\left(2\mathcal{E}_{ij}\nabla^2-\mathcal{F}_{ij}r\frac{\partial}{\partial
r}\frac{1}{r}\frac{\partial}{\partial r}\right)Y(\Lambda,m_a,r),
\end{eqnarray}
where $\mathcal{D}_{ij},\mathcal{E}_{ij}$, and $\mathcal{F}_{ij}$
are the spin-spin interactions and tensor force operators, their
expressions are defined as follows, e.g.,
\begin{eqnarray*}
\left.\begin{array}{ll} \mathcal{D}_{12} =
\sum_{a,b}C_{\frac{1}{2},m;1,n}^{\frac{3}{2},m+n}\chi_{3,m}^{\dag}
    \bm{\epsilon}_{3,n}^{\dag}\cdot\bm{\sigma}\chi_1,\\
\mathcal{E}_{13} =
\bm{\epsilon}_4^{\dag}\cdot\bm{\sigma},\quad\quad\quad\quad
\mathcal{F}_{13} = S\left(\hat{r},\bm{\epsilon}_4^{\dag},\bm{\sigma}\right),\\
\mathcal{E}_{14} =
\sum_{m,n}C_{\frac{1}{2},m;1,n}^{\frac{3}{2},m+n}\chi_{3,m}^{\dag}\bm{\epsilon}_4^{\dag}\cdot
\left(i\bm{\sigma}\times\bm{\epsilon}_{3,n}^{\dag}\right)\chi_1,\\
\mathcal{F}_{14} =
\sum_{m,n}C_{\frac{1}{2},m;1,n}^{\frac{3}{2},m+n}\chi_{3,m}^{\dag}
  S\left(\hat{r},\bm{\epsilon}_4^{\dag},i\bm{\sigma}\times\bm{\epsilon}_{3,n}^{\dag}\right)\chi_1,\\
\mathcal{D}_{22} =
\sum_{a,b}^{m,n}C_{\frac{1}{2},a;1,b}^{\frac{3}{2},a+b}
   C_{\frac{1}{2},m;1,n}^{\frac{3}{2},m+n}\chi_3^{a\dag}\chi_1^m
   \bm{\epsilon}_1^n\cdot\bm{\epsilon}_3^{b\dag},\\
\mathcal{E}_{23} =
     \sum_{m,n}C_{\frac{1}{2},m;1,n}^{\frac{3}{2},m+n}\chi_3^{m\dag}
     \bm{\epsilon}_2\cdot\left(i\bm{\epsilon}_3^{n\dag}\times\bm{\sigma}\right)\chi_1,\\
\mathcal{F}_{23} =
\sum_{m,n}C_{\frac{1}{2},m;1,n}^{\frac{3}{2},m+n}\chi_3^{m\dag}
     S\left(\hat{r},\bm{\epsilon}_2,i\bm{\epsilon}_3^{n\dag}\times\bm{\sigma}\right)\chi_1,\\
\mathcal{E}_{24} =
\sum_{a,b}^{m,n}C_{\frac{1}{2},a;1,b}^{\frac{3}{2},a+b}
     C_{\frac{1}{2},m;1,n}^{\frac{3}{2},m+n}\chi_3^{a\dag}\chi_1^{m}
     \bm{\epsilon}_2\cdot\left(i\bm{\epsilon}_1^{n}\times
                 \bm{\epsilon}_3^{b\dag}\right),\\
\mathcal{F}_{24} =
\sum_{a,b}^{m,n}C_{\frac{1}{2},a;1,b}^{\frac{3}{2},a+b}
     C_{\frac{1}{2},m;1,n}^{\frac{3}{2},m+n}\chi_3^{a\dag}\chi_1^{m}
                 S\left(\hat{r},\bm{\epsilon}_2,i\bm{\epsilon}_1^{n}
                 \times\bm{\epsilon}_3^{b\dag}\right),\\
\mathcal{D}_{33} =
\chi_3^{\dag}\chi_1\bm{\epsilon}_2\cdot\bm{\epsilon}_4^{\dag},\quad\quad\quad\quad
\mathcal{E}_{33} =
\chi_3^{\dag}\bm{\sigma}\cdot\left(i\bm{\epsilon}_2\times
     \bm{\epsilon}_4^{\dag}\right)\chi_1,\\
\mathcal{F}_{33} = \chi_3^{\dag}S\left(\hat{r},\bm{\sigma},
     i\bm{\epsilon}_2\times\bm{\epsilon}_4^{\dag}\right)\chi_1,\\
\mathcal{D}_{34} =
\sum_{m,n}C_{\frac{1}{2},m;1,n}^{\frac{3}{2},m+n}\chi_3^{\dag}
   \left(\bm{\sigma}\cdot\bm{\epsilon}_1\right)\left(\bm{\epsilon}_2\cdot\bm{\epsilon}_4^{\dag}
   \right)\chi_1^m,\\
\mathcal{E}_{34} =
\sum_{m,n}C_{\frac{1}{2},m;1,n}^{\frac{3}{2},m+n}\chi_3^{\dag}
    \left(\bm{\sigma}\times\bm{\epsilon}_1^{n}\right)\cdot
    \left(\bm{\epsilon}_2\times\bm{\epsilon}_4^{\dag}\right)\chi_1^m,\\
\mathcal{F}_{34} =
\sum_{m,n}C_{\frac{1}{2},m;1,n}^{\frac{3}{2},m+n}\chi_3^{\dag}
    S\left(\hat{r},\bm{\sigma}\times\bm{\epsilon}_1,\bm{\epsilon}_2\times\bm{\epsilon}_4^{\dag}\right)
    \chi_1^m,\\
\mathcal{D}_{44} =
\sum_{a,b}^{m,n}C_{\frac{1}{2},a;1,b}^{\frac{3}{2},a+b}
    C_{\frac{1}{2},m;1,n}^{\frac{3}{2},m+n}\chi_3^{a\dag}
    \left(\bm{\epsilon}_1^n\cdot\bm{\epsilon}_3^{b\dag}\right)
    \left(\bm{\epsilon}_2\cdot\bm{\epsilon}_4^{\dag}\right)\chi_1^m,\\
\mathcal{E}_{44} =
\sum_{a,b}^{m,n}C_{\frac{1}{2},a;1,b}^{\frac{3}{2},a+b}
    C_{\frac{1}{2},m;1,n}^{\frac{3}{2},m+n}\chi_3^{a\dag}
    \left(\bm{\epsilon}_1^n\times\bm{\epsilon}_3^{b\dag}\right)
    \cdot\left(\bm{\epsilon}_2\times\bm{\epsilon}_4^{\dag}\right)\chi_1^m,\\
\mathcal{F}_{44} =
\sum_{a,b}^{m,n}C_{\frac{1}{2},a;1,b}^{\frac{3}{2},a+b}
    C_{\frac{1}{2},m;1,n}^{\frac{3}{2},m+n}\chi_3^{a\dag}
    S\left(\hat{r},\bm{\epsilon}_1^n\times\bm{\epsilon}_3^{b\dag},
    \bm{\epsilon}_2\times\bm{\epsilon}_4^{\dag}\right)\chi_1^m.
\end{array}\right.\nonumber
\end{eqnarray*}
Once we start the numerical calculation through solving the coupled
channel Shr$\ddot{\text{o}}$dinger equation, these operators
$\mathcal{O}_{ij}$ should be replaced by a serial of matrix elements
$\langle{}^{2s'+1}L'_{J'}|\mathcal{O}_{ij}|{}^{2s+1}L_J\rangle$ as
collected in Table \ref{potentials}, where the notations
$\langle{}^{2s'+1}L'_{J'}|$ and $|{}^{2s+1}L_J\rangle$ stand for the
spin-orbit wave functions for the final and initial discussed
channels, respectively.

\renewcommand\tabcolsep{0.2cm}
\renewcommand{\arraystretch}{1.8}
\begin{table*}[!htbp]
\caption{The OBE effective potentials for the
$\Sigma_c^{(*)}{D}^{(*)}\to\Sigma_c^{(*)}{D}^{(*)}$ processes and
the matrix elements $\langle \mathcal{O}_{ij}\rangle$ obtained from
$\langle{}^{2s'+1}L'_{J'}|\mathcal{O}_{ij}|{}^{2s+1}L_J\rangle$ for
all the operators $\mathcal{O}_{ij}$. Here, $\mathcal{G}$ is the
isospin factor, which is taken as $-1$ for the isospin-$1/2$ system,
and ${1}/{2}$ for the isospin-$3/2$ system. The values of the
coupling constants are taken from
\cite{Liu:2007bf,Liu:2011xc,Casalbuoni:1996pg,Falk:1992cx},
$g_S=0.76,g=0.59,\beta=0.9,l_S=6.2,g_1=0.94,\beta_S=-1.74,$
$\lambda=0.56$ GeV$^{-1}$, $\lambda_S=-3.31$ GeV$^{-1}$, and
$g_V=5.9$. The variables in these functions are defined as
$\Lambda_i^2 =\Lambda^2-q_i^2$, $m_{{i}}^2=m^2-q_i^2$, with $i=0$,
1, ..., 12, and $q_i^2 =
\left(\frac{M_{A}^2+M_{D}^2-M_{C}^2-M_{B}^2}{2(M_{C}+M_{D})}\right)^2$
for the $A+B\to C+D$ process.
$\langle\mathcal{D}_{12}\rangle=\langle\mathcal{D}_{34}\rangle=(0)$,
$\langle\mathcal{D}_{22}\rangle=(\bf{1})$,
$\langle\mathcal{E}_{13}\rangle(1/2^-)=(\sqrt{3}, 0)^T$,
$\langle\mathcal{F}_{13}\rangle(1/2^-)=(0, -\sqrt{6})^T$,
$\langle\mathcal{E}_{14}\rangle(1/2^-)=(\sqrt{2}, 0, 0)^T$,
$\langle\mathcal{F}_{14}\rangle(1/2^-)=(0, \sqrt{\frac{2}{5}},
-3\sqrt{\frac{2}{5}})^T$.}\label{potentials}
{\begin{tabular}{c|ll|c|ll} \toprule[1pt]\toprule[1pt]
  \multicolumn{2}{c}{Processes}   &\multicolumn{4}{l}{OBE effective potentials}\\
\midrule[1pt]
 \multicolumn{2}{c}{$\Sigma_c{D}\to\Sigma_c{D}$}
    &\multicolumn{4}{l}{$-l_Sg_SY(\Lambda,m_{\sigma},r)
       -\frac{\mathcal{G}\beta\beta_Sg_V^2}{2}Y(\Lambda,m_{\rho},r)
       +\frac{\beta\beta_Sg_V^2}{4}Y(\Lambda,m_{\omega},r)$}\\
 \multicolumn{2}{c}{$\Sigma_c{D}\to\Sigma_c^{*}{D}$}      &\multicolumn{4}{l}{$\frac{\mathcal{G}\beta\beta_Sg_V^2}{2\sqrt{3}}
       \mathcal{Y}^{12}_{\Lambda_3,m_{\rho3}}
       -\frac{\beta\beta_Sg_V^2}{4\sqrt{3}}\mathcal{Y}^{12}_{\Lambda_3,m_{\omega3}}$}\\
 \multicolumn{2}{c}{$\Sigma_c{D}\to\Sigma_c{D}^{*}$}      &\multicolumn{4}{l}{$-\frac{\mathcal{G}}{3}\frac{gg_1}{f_\pi^2}\mathcal{Z}^{13}_{\Lambda_4,m_{\pi4}}
        +\frac{1}{18}\frac{gg_1}{f_\pi^2}\mathcal{Z}^{13}_{\Lambda_4,m_{\eta4}}
        -\frac{2\mathcal{G}\lambda\lambda_Sg_V^2}{9}\mathcal{Z}^{\prime13}_{\Lambda_4,m_{\rho4}}
        +\frac{\lambda\lambda_Sg_V^2}{9}\mathcal{Z}^{\prime13}_{\Lambda_4,m_{\omega4}}$}\\
 \multicolumn{2}{c}{$\Sigma_c{D}\to\Sigma_c^{*}{D}^{*}$}   &\multicolumn{4}{l}{$-\frac{\mathcal{G}}{2\sqrt{3}}\frac{gg_1}{f_\pi^2}
       \mathcal{Z}^{14}_{\Lambda_5,m_{\pi5}}
        +\frac{1}{12\sqrt{3}}\frac{gg_1}{f_\pi^2}\mathcal{Z}^{14}_{\Lambda_5,m_{\eta5}}
        +\frac{\mathcal{G}\lambda\lambda_Sg_V^2}{3\sqrt{3}}\mathcal{Z}^{\prime14}_{\Lambda_5,m_{\rho5}}
        -\frac{\lambda\lambda_Sg_V^2}{6\sqrt{3}}\mathcal{Z}^{\prime14}_{\Lambda_5,m_{\omega5}}$}\\
 \multicolumn{2}{c}{$\Sigma_c^{*}{D}\to\Sigma_c^{*}{D}$}   &\multicolumn{4}{l}{$-l_Sg_S\mathcal{Y}^{22}_{\Lambda,m_{\sigma}}
              -\frac{\mathcal{G}\beta\beta_Sg_V^2}{2}\mathcal{Y}^{22}_{\Lambda,m_{\rho}}
                 +\frac{\beta\beta_Sg_V^2}{4}\mathcal{Y}^{22}_{\Lambda,m_{\omega}}$}\\
 \multicolumn{2}{c}{$\Sigma_c^{*}{D}\to\Sigma_c{D}^*$}   &\multicolumn{4}{l}{$-\frac{\mathcal{G}}{2\sqrt{3}}\frac{gg_1}{f_\pi^2}
     \mathcal{Z}^{23}_{\Lambda_0,m_{\pi0}}
    +\frac{1}{12\sqrt{3}}\frac{gg_1}{f_\pi^2}\mathcal{Z}^{23}_{\Lambda_0,m_{\eta0}}
    +\frac{\mathcal{G}\lambda\lambda_Sg_V^2}{3\sqrt{3}}\mathcal{Z}^{\prime23}_{\Lambda_0,m_{\rho0}}
    -\frac{\lambda\lambda_Sg_V^2}{6\sqrt{3}}\mathcal{Z}^{\prime23}_{\Lambda_0,m_{\omega0}}$}\\
 \multicolumn{2}{c}{$\Sigma_c^{*}{D}\to\Sigma_c^{*}{D}^{*}$}   &\multicolumn{4}{l}{$\frac{\mathcal{G}}{2}\frac{gg_1}{f_\pi^2}\mathcal{Z}^{24}_{\Lambda_1,m_{\pi1}}
    -\frac{1}{12}\frac{gg_1}{f_\pi^2}\mathcal{Z}^{24}_{\Lambda_1,m_{\eta1}}
    +\frac{\mathcal{G}\lambda\lambda_Sg_V^2}{3}\mathcal{Z}^{\prime24}_{\Lambda_1,m_{\rho1}}
    -\frac{\lambda\lambda_Sg_V^2}{6}\mathcal{Z}^{\prime 24}_{\Lambda_1,m_{\omega1}}$}\\
 \multicolumn{2}{c}{$\Sigma_c{D}^{*}\to\Sigma_c{D}^{*}$}      &\multicolumn{4}{l}{$-l_Sg_S\mathcal{Y}^{33}_{\Lambda,m_{\sigma}}
     -\frac{\mathcal{G}}{3}\frac{gg_1}{f_\pi^2}\mathcal{Z}^{33}_{\Lambda,m_{\pi}}
     +\frac{1}{18}\frac{gg_1}{f_\pi^2}\mathcal{Z}^{33}_{\Lambda,m_{\eta}}
    -\frac{\mathcal{G}\beta\beta_Sg_V^2}{2}\mathcal{Y}^{33}_{\Lambda,m_{\rho}}
    -\frac{2\mathcal{G}\lambda\lambda_Sg_V^2}{9}\mathcal{Z}^{\prime33}_{\Lambda,m_{\rho}}
    -\frac{\beta\beta_Sg_V^2}{4}\mathcal{Y}^{33}_{\Lambda,m_{\omega}}
    +\frac{\lambda\lambda_Sg_V^2}{9}\mathcal{Z}^{\prime33}_{\Lambda,m_{\omega}}$}\\
 \multicolumn{2}{c}{$\Sigma_c{D}^{*}\to\Sigma_c^{*}{D}^{*}$}   &\multicolumn{4}{l}{$\frac{l_Sg_S}{\sqrt{3}}\mathcal{Y}^{34}_{\Lambda_2,m_{\sigma2}}
    -\frac{\sqrt{3}\mathcal{G}}{6}\frac{gg_1}{f_\pi^2}\mathcal{Z}^{34}_{\Lambda_2,m_{\pi2}}
    +\frac{\sqrt{3}}{36}\frac{gg_1}{f_\pi^2}\mathcal{Z}^{34}_{\Lambda_2,m_{\eta2}}
        +\frac{\mathcal{G}\beta\beta_Sg_V^2}{2\sqrt{3}}\mathcal{Y}^{34}_{\Lambda_2,m_{\rho2}}
        -\frac{\mathcal{G}\lambda\lambda_Sg_V^2}{3\sqrt{3}}\mathcal{Z}^{\prime34}_{\Lambda_2,m_{\rho2}}
        +\frac{\beta\beta_Sg_V^2}{4\sqrt{3}}\mathcal{Y}^{34}_{\Lambda_2,m_{\omega2}}
        +\frac{\lambda\lambda_Sg_V^2}{6\sqrt{3}}\mathcal{Z}^{\prime 34}_{\Lambda_2,m_{\omega2}}$}\\
 \multicolumn{2}{c}{$\Sigma_c^{*}{D}^{*}\to\Sigma_c^{*}{D}^{*}$}   &\multicolumn{4}{l}{$-l_Sg_S\mathcal{Y}^{44}_{\Lambda,m_{\sigma}}
       +\frac{\mathcal{G}}{2}\frac{gg_1}{f_\pi^2}\mathcal{Z}^{44}_{\Lambda,m_{\pi}}
       -\frac{1}{12}\frac{gg_1}{f_\pi^2}\mathcal{Z}^{44}_{\Lambda,m_{\eta}}
       -\frac{\mathcal{G}\beta\beta_Sg_V^2}{2}\mathcal{Y}^{44}_{\Lambda,m_{\rho}}
       +\frac{\mathcal{G}\lambda\lambda_Sg_V^2}{3}\mathcal{Z}^{\prime44}_{\Lambda,m_{\rho}}
       +\frac{\beta\beta_Sg_V^2}{4}\mathcal{Y}^{44}_{\Lambda,m_{\omega}}
       -\frac{\lambda\lambda_Sg_V^2}{6}\mathcal{Z}^{\prime44}_{\Lambda,m_{\omega}}$}\\
\bottomrule[1pt]\bottomrule[1pt] $\langle\mathcal{O}_{ij}\rangle$
&$J^P=1/2^-$    &$J^P=3/2^-$
  &$\langle\mathcal{O}_{ij}\rangle$     &$J^P=1/2^-$   &$J^P=3/2^-$    \\\hline
  $\mathcal{E}_{23}$     &$\left(\begin{array}{cc} 0 & 1\end{array}\right)$
           &$\left(\begin{array}{ccc} 1 & 0 & 0 \\ 0 & 0 & 1\end{array}\right)$
  &$\mathcal{E}_{44}$
            &{$\left(\begin{array}{ccc} \frac{5}{3} & 0 & 0 \\ 0 & \frac{2}{3} & 0 \\ 0 & 0 & -1\end{array}\right)$}
            &{$\left(\begin{array}{ccc} \frac{5}{3} & 0 & 0 \\ 0 & \frac{2}{3} & 0 \\ 0 & 0 & -1\end{array}\right)$}\\
  $\mathcal{F}_{23}$      &$\left(\begin{array}{cc} -\sqrt{2} & 1\end{array}\right)$
           &$\left(\begin{array}{ccccc} 0 & 1 & -1 \\ -1 & -1 & 0\end{array}\right)$
  &$\mathcal{E}_{24}$     &{$\left(\begin{array}{ccc} 0 & -\frac{1}{3}\sqrt{\frac{5}{3}} & 0\end{array}\right)$}
            &{$\left(\begin{array}{cccc} \frac{1}{\sqrt{15}} & 0 & 0 & 0 \\ 0 & -\frac{2 }{15}\sqrt{\frac{2}{3}} & -\frac{49}{15 \sqrt{15}} & -\frac{2 }{5}\sqrt{\frac{14}{15}}\end{array}\right)$}\\
  $\mathcal{D}_{33}$
            &$\left(\begin{array}{cc} 1 & 0 \\ 0 & 1\end{array}\right)$
            &{$\left(\begin{array}{ccc} 1 & 0 & 0 \\ 0 & 1 & 0 \\ 0 & 0 & 1\end{array}\right)$}
  &$\mathcal{F}_{24}$    &{$\left(\begin{array}{ccc} -\frac{4}{15 \sqrt{3}} & \frac{2}{3 \sqrt{15}} & 0\end{array}\right)$}
            &{$\left(\begin{array}{cccc} 0 & \frac{1}{5}\sqrt{\frac{3}{2}} & -\frac{8}{5 \sqrt{15}} & -\frac{1}{5}\sqrt{\frac{3}{70}} \\ -\frac{2}{5 \sqrt{15}} & \frac{17}{105 \sqrt{6}} & -\frac{22}{105 \sqrt{15}} & \frac{1}{35 \sqrt{210}}\end{array}\right)$}\\
  $\mathcal{E}_{33}$
            &$\left(\begin{array}{cc} -2 & 0 \\ 0 & 1\end{array}\right)$
            &{$\left(\begin{array}{ccc} 1 & 0 & 0 \\ 0 & -2 & 0 \\ 0 & 0 & 1\end{array}\right)$}
  &$\mathcal{E}_{34}$
            &$\left(\begin{array}{ccc} \frac{\sqrt{\frac{2}{3}}}{3} & 0 &0\\ 0 & \frac{\sqrt{\frac{5}{3}}}{3}  & 0\end{array}\right)$
            &{$\left(\begin{array}{ccc} 1 & 0 & 0 \\ 0 & -2 & 0 \\ 0 & 0 & 1\end{array}\right)$}\\
  $\mathcal{F}_{33}$
            &$\left(\begin{array}{cc} 0 & -\sqrt{2} \\ -\sqrt{2} & -2\end{array}\right)$
            &{$\left(\begin{array}{ccc} 0 & 1 & 2 \\ 1 & 0 & -1 \\ 2 & -1 & 0\end{array}\right)$}
  &$\mathcal{F}_{34}$
            &{$\left(\begin{array}{ccc} 0   &-\frac{4 \sqrt{\frac{2}{15}}}{15}   &-\frac{11 \sqrt{\frac{2}{15}}}{5}\\  \frac{17}{15 \sqrt{3}}   & \frac{19}{15 \sqrt{15}}   & \frac{1}{5 \sqrt{15}}\end{array}\right)$}
             &{$\left(\begin{array}{cccc} 0   &-\frac{\sqrt{\frac{2}{3}}}{5}   &-\frac{11}{5 \sqrt{15}}   &-\frac{\sqrt{\frac{6}{35}}}{5}\\  -\frac{2}{5 \sqrt{15}}  & \frac{\sqrt{\frac{2}{3}}}{15} & -\frac{94}{105 \sqrt{15}}  & \frac{46 \sqrt{\frac{2}{105}}}{35}\\   -\frac{1}{\sqrt{15}}  & \frac{76 \sqrt{\frac{2}{3}}}{105}   & -\frac{46}{105 \sqrt{15}}  & \frac{64 \sqrt{\frac{2}{105}}}{35}\end{array}\right)$}\\
  $\mathcal{D}_{44}$
            &{$\left(\begin{array}{ccc} 1 & 0 & 0 \\ 0 & 1 & 0 \\ 0 & 0 & 1\end{array}\right)$}
            &{$\left(\begin{array}{cccc} 1 & 0 & 0 & 0 \\ 0 & 1 & 0 & 0 \\ 0 & 0 & 1 & 0 \\ 0 & 0 & 0 & 1\end{array}\right)$}
  &$\mathcal{F}_{44}$
            &{$\left(\begin{array}{ccc} 0 & -\frac{7}{3 \sqrt{5}} & \frac{2}{\sqrt{5}} \\ -\frac{7}{3 \sqrt{5}} & \frac{16}{15} & -\frac{1}{5} \\ \frac{2}{\sqrt{5}} & -\frac{1}{5} & \frac{8}{5}\end{array}\right)$}
            &{$\left(\begin{array}{cccc} 0 & \frac{7}{3\sqrt{10}} & -\frac{16}{15} & -\frac{\sqrt{\frac{7}{2}}}{5} \\ \frac{7}{3\sqrt{10}} & 0 & -\frac{7}{3\sqrt{10}} & -\frac{2}{\sqrt{35}} \\ -\frac{16}{15} & -\frac{7}{3\sqrt{10}} & 0 & -\frac{1}{\sqrt{14}} \\ -\frac{\sqrt{\frac{7}{2}}}{5} & -\frac{2}{\sqrt{35}} & -\frac{1}{\sqrt{14}} & \frac{4}{7}\end{array}\right)$}\\
   \bottomrule[2pt]
\end{tabular}}
\end{table*}

\section{Numerical results}\label{sec3}

We perform a systematic investigation on the possible molecular
pentaquarks composed of the $S-$wave $\Sigma_c^{(*)}D^{(*)}$ systems
with all possible isospin $I$ and spin $J$ for the negative parity.
To explore the roles of the $S-D$ wave mixing effects, the coupled
channel effects, the long-rang pion-exchange potential and the
intermediate- and short-range from the $\rho$, $\omega$, $\sigma$,
and $\eta$ exchanges in the formation of the loosely bound
$\Sigma_c^{(*)}D^{(*)}$ states, we first perform the calculation for
the single channel with both the one-pion-exchange (OPE) potentials
and OBE potentials, the numerical results of which are given in
Table~\ref{num1}. We then include the coupled channel effects and
perform the calculation again using the OPE potentials, and present
the numerical results in Table~\ref{num2}. Finally, we include the
intermediate- and short-range $\rho$, $\omega$, $\sigma$ and $\eta$
exchanges potentials in addition to the long-range interaction
potentials, and the numerical results are presented in
Table~\ref{num3}.

\renewcommand\tabcolsep{0.11cm}
\renewcommand{\arraystretch}{1.8}
\begin{table}[!htbp]
\caption{Bound state solutions for the $S-$wave single
$\Sigma_c^{(*)}D^{(*)}$ systems. The cutoff $\Lambda$, the
root-mean-square radius $r_{RMS}$, and the binding energy $E$ of the
bound state are in units of GeV, fm, and MeV,
respectively.}\label{num1}
\begin{tabular}{c|ccc|ccc}
\toprule[2pt] &\multicolumn{3}{c|}{OPE}
&\multicolumn{3}{c}{OBE}\\\hline
&$\Lambda$    &$E$    &$r_{RMS}$    &$\Lambda$    &$E$    &$r_{RMS}$\\
$\Sigma_c{D}[1/2(1/2^-)]$    &\ldots    &\ldots     &\ldots   &1.14      &$-0.23$       &5.21\\
                                 &\ldots    &\ldots     &\ldots   &1.24      &$-5.95$       &1.53\\
                                 &\ldots    &\ldots     &\ldots   &1.34      &$-18.18$      &0.97\\\hline
$\Sigma_c{D}[3/2(1/2^-)]$    &\ldots    &\ldots     &\ldots
&\ldots    &\ldots     &\ldots\\\hline
$\Sigma_c^*{D}[1/2(3/2^-)]$      &\ldots    &\ldots     &\ldots     &1.15     &$-0.57$     &4.05\\
                                 &\ldots    &\ldots     &\ldots     &1.25     &$-7.30$     &1.40\\
                                 &\ldots    &\ldots     &\ldots     &1.35     &$-20.42$    &0.92\\\hline
$\Sigma_c^*{D}[3/2(3/2^-)]$      &\ldots    &\ldots     &\ldots
&\ldots    &\ldots     &\ldots\\\hline
$\Sigma_c{D}^*[1/2(1/2^-)]$  &2.40      &$-0.23$        &5.28        &1.30    &$-0.51$    &4.41\\
                                 &2.55      &$-3.04$        &2.03        &1.40    &$-3.75$    &1.99\\
                                 &2.70      &$-10.90$        &1.18       &1.50    &$-10.47$   &1.33\\\hline
$\Sigma_c{D}^*[3/2(1/2^-)]$  &1.64      &$-0.35$        &4.43        &1.45    &$-0.69$    &3.60\\
                                 &1.80      &$-5.75$        &1.33        &1.65    &$-5.50$    &1.43\\
                                 &1.96      &$-18.81$        &0.78       &1.85    &$-15.27$   &0.91\\\hline
$\Sigma_c{D}^*[1/2(3/2^-)]$  &1.20    &$-1.30$    &2.86     &0.91     &$-0.30$     &4.85\\
                                 &1.30    &$-4.99$    &1.60     &0.96     &$-3.30$     &1.96\\
                                 &1.40    &$-11.61$   &1.13     &1.01     &$-9.74$     &1.26\\\hline
$\Sigma_c{D}^*[3/2(3/2^-)]$  &4.00      &$-0.56$        &3.98      &3.90    &$-0.42$    &4.42\\
                                 &4.30      &$-4.95$        &1.54      &4.40    &$-3.68$    &1.79\\
                                 &4.60      &$-15.36$        &0.93     &4.90    &$-11.83$   &1.06\\
\bottomrule[2pt]
\end{tabular}
\end{table}

\renewcommand\tabcolsep{0.11cm}
\renewcommand{\arraystretch}{1.8}
\begin{table*}[!htbp]
\caption{Bound state solutions for the coupled
$\Sigma_c^{(*)}D^{(*)}$ systems using the OPE potential. The cutoff
$\Lambda$, the root-mean-square radius $r_{RMS}$, and binding energy
$E$ of the bound state are in units of GeV, fm, and MeV,
respectively. $P_i(\%)$ denotes the probability of the $i-$th
channel. The results for the channel with the largest probability
are marked with a bold typeface.}\label{num2}
\begin{tabular}{cc|cccc}
\toprule[2pt] \multicolumn{2}{c|}{OPE}  &\multicolumn{4}{c}{Channels
($P_i$)}\\\hline
$I(J^P)$    &[$\Lambda$, $E$, $r_{RMS}$]   &$\Sigma_cD|{}^2S_{1/2}\rangle$    &$\Sigma_c^*D|{}^4D_{1/2}\rangle$    &$\Sigma_cD^*|{}^2S_{1/2}/{}^4D_{1/2}\rangle$        &$\Sigma_c^*D^*|{}^2S_{1/2}/{}^4D_{1/2}/{}^6D_{1/2}\rangle$\\
$1/2(1/2^-)$    &$[1.21, -0.87, 3.41]$     &\textbf{96.90}  &$\sim 0$    &1.32/0.77   &0.39/0.06/0.55\\
                &$[1.29, -6.15, 1.43]$     &\textbf{91.72}  &0.01        &3.72/1.75   &1.14/0.17/1.48\\
                &$[1.37, -16.93, 0.93]$    &\textbf{86.63}  &0.02        &6.20/2.44   &1.97/0.28/2.44\\\hline
$3/2(1/2^-)$    &$[1.62, -0.64, 3.65]$     &\textbf{93.86}   &0.01        &3.85/0.43   &1.78/$\sim 0$/0.06\\
                &$[1.67, -5.40, 1.31]$     &\textbf{82.12}   &0.03        &11.45/0.83   &5.47/$\sim 0$/0.09\\
                &$[1.72, -14.88, 0.79]$    &\textbf{71.46}   &0.05        &18.31/0.92   &9.16/$\sim 0$/0.08\\\midrule[1pt]
$1/2(1/2^-)$    &$[2.24, -0.52, 4.25]$    &\ldots  &\ldots    &\textbf{93.85}/4.35    &0.39/0.12/1.28\\
                &$[2.34, -3.83, 1.78]$    &\ldots  &\ldots    &\textbf{78.81}/8.04    &4.80/1.25/7.10\\
                &$[2.44, -15.89, 0.85]$   &\ldots  &\ldots    &\textbf{41.17}/4.28    &25.16/6.30/23.09\\\hline
$3/2(1/2^-)$    &$[1.58, -0.36,4.39]$     &\ldots  &\ldots    &\textbf{98.67}/0.40    &0.90/0.01/0.02\\
                &$[1.68, -3.97, 1.54]$    &\ldots  &\ldots    &\textbf{93.31}/0.83    &5.77/0.07/0.02\\
                &$[1.78, -13.40, 0.84]$   &\ldots  &\ldots    &\textbf{78.58}/0.98    &20.20/0.23/$\sim 0$\\\midrule[1pt]
&&&$\Sigma_c^*D|{}^4S_{3/2}/{}^4D_{3/2}\rangle$ &$\Sigma_cD^*|{}^4S_{3/2}/{}^2D_{3/2}/{}^4D_{3/2}\rangle$        &$\Sigma_c^*D^*|{}^4S_{3/2}/{}^2D_{3/2}/{}^4D_{3/2}/{}^6D_{3/2}\rangle$\\
$1/2(3/2^-)$    &$[1.48, -0.76, 3.09]$   &\ldots    &\textbf{77.34}/0.13    &20.95/0.84/0.36     &0.25/0.08/0.04/$\sim 0$\\
                &$[1.50, -3.22, 1.44]$   &\ldots    &\textbf{59.44}/0.24    &\textbf{37.53}/1.33/0.80     &0.48/0.14/0.05/$\sim 0$\\
                &$[1.52, -6.82, 0.96]$   &\ldots    &\textbf{48.10}/0.31    &\textbf{48.01}/1.56/1.17     &0.64/0.16/0.04/$\sim 0$\\\hline
$3/2(3/2^-)$    &$[2.39, -0.64, 3.49]$   &\ldots    &\textbf{87.05}/0.06    &4.63/0.060.84       &6.67/0.14/0.52/0.02\\
                &$[2.44, -5.25, 1.20]$   &\ldots    &\textbf{67.54}/0.16    &11.24/0.07/1.85     &17.48/0.33/1.26/0.05\\
                &$[2.49, -13.29, 0.75]$   &\ldots   &\textbf{54.82}/0.22    &15.06/0.06/2.33     &25.26/0.46/1.73/0.07\\\midrule[1pt]
&&&&$\Sigma_cD^*|{}^4S_{3/2}/{}^2D_{3/2}/{}^4D_{3/2}\rangle$        &$\Sigma_c^*D^*|{}^4S_{3/2}/{}^2D_{3/2}/{}^4D_{3/2}/{}^6D_{3/2}\rangle$\\
$1/2(3/2^-)$    &$[1.12, -0.51, 4.12]$   &\ldots &\ldots &\textbf{96.79}/0.70/2.31   &0.16/0.01/0.02/$\sim 0$\\
                &$[1.22, -3.42, 1.87]$   &\ldots &\ldots &\textbf{93.97}/1.28/4.22   &0.47/0.03/0.03/$\sim 0$\\
                &$[1.32, -9.42, 1.22]$   &\ldots &\ldots &\textbf{92.08}/1.60/5.37   &0.85/0.05/0.03/$\sim 0$\\\hline
$3/2(3/2^-)$    &$[2.54, -0.72, 3.32]$   &\ldots &\ldots &\textbf{81.69}/0.32/1.71   &15.24/0.14/0.85/0.05\\
                &$[2.60, -5.31, 1.17]$   &\ldots &\ldots &\textbf{55.43}/0.47/2.53   &\textbf{38.99}/0.39/2.05/0.14\\
                &$[2.66, -13.26, 0.73]$  &\ldots &\ldots &\textbf{40.42}/0.46/2.46   &\textbf{53.16}/0.57/2.73/0.19\\
\bottomrule[2pt]
\end{tabular}
\end{table*}

\renewcommand\tabcolsep{0.11cm}
\renewcommand{\arraystretch}{1.8}
\begin{table*}[!htbp]
\caption{Bound state solutions for the coupled
$\Sigma_c^{(*)}D^{(*)}$ systems using the OBE potential. The cutoff
$\Lambda$, the root-mean-square radius $r_{RMS}$, and the binding
energy $E$ of the bound state are in units of GeV, fm, and MeV,
respectively. $P_i(\%)$ denotes the probability of the $i-$th
channel. The results for the channel with the largest probability
are marked with a bold typeface.}\label{num3}
\begin{tabular}{cc|cccc}
\toprule[2pt] \multicolumn{2}{c|}{OBE}  &\multicolumn{4}{c}{Channels
($p_i$)}\\\hline
$I(J^P)$    &[$\Lambda$, $E$, $r_{RMS}$]   &$\Sigma_cD|{}^2S_{1/2}\rangle$    &$\Sigma_c^*D|{}^4D_{1/2}\rangle$    &$\Sigma_cD^*|{}^2S_{1/2}/{}^4D_{1/2}\rangle$        &$\Sigma_c^*D^*|{}^2S_{1/2}/{}^4D_{1/2}/{}^6D_{1/2}\rangle$\\
$1/2(1/2^-)$    &$[0.96, -0.49, 4.29]$     &\textbf{98.62}  &$\sim 0$  &0.61/0.37   &0.18/0.02/0.20\\
                &$[1.01, -5.31, 1.59]$     &\textbf{95.42}  &$\sim 0$  &2.20/1.00   &0.69/0.07/0.62\\
                &$[1.06, -15.47, 1.03]$    &\textbf{92.18}  &$\sim 0$  &3.89/1.43   &1.32/0.12/1.04\\\hline
$3/2(1/2^-)$    &$[1.75, -0.53, 4.05]$     &\textbf{97.78}  &$\sim 0$  &0.50/0.34   &1.32/$\sim 0$/0.05\\
                &$[1.85, -4.24, 1.61]$     &\textbf{92.02}  &$\sim 0$  &1.49/0.85   &5.53/$\sim 0$/0.10\\
                &$[1.95, -13.52, 0.92]$    &\textbf{82.43}  &$\sim 0$  &2.59/1.24   &13.62/$\sim 0$/0.10\\\midrule[1pt]
$1/2(1/2^-)$    &$[1.27, -0.41, 4.696]$    &\ldots  &\ldots    &\textbf{97.56}/2.24   &0.01/$\sim 0$/0.19\\
                &$[1.37, -3.79, 1.99]$     &\ldots  &\ldots    &\textbf{93.07}/6.09   &0.07/$\sim 0$/0.76\\
                &$[1.47, -11.22, 1.30]$    &\ldots  &\ldots    &\textbf{88.46}/9.44   &0.30/0.03/1.76\\\hline
$3/2(1/2^-)$    &$[1.41, -0.48, 4.09]$     &\ldots  &\ldots    &\textbf{99.53}/0.28   &0.17/$\sim 0$/0.02\\
                &$[1.61, -5.45, 1.43]$     &\ldots  &\ldots    &\textbf{97.85}/0.56   &1.53/0.01/0.03\\
                &$[1.81, -17.34, 0.84]$    &\ldots  &\ldots    &\textbf{91.13}/0.73   &8.04/0.07/0.02\\\midrule[1pt]
&&&$\Sigma_c^*D|{}^4S_{3/2}/{}^4D_{3/2}\rangle$ &$\Sigma_cD^*|{}^4S_{3/2}/{}^2D_{3/2}/{}^4D_{3/2}\rangle$        &$\Sigma_c^*D^*|{}^4S_{3/2}/{}^2D_{3/2}/{}^4D_{3/2}/{}^6D_{3/2}\rangle$\\
$1/2(3/2^-)$    &$[1.05, -1.30, 2.84]$    &\ldots    &\textbf{94.50}/0.02     &5.14/0.27/0.02    &$\sim 0$/0.02/0.03/$\sim 0$\\
                &$[1.08, -5.24, 1.48]$    &\ldots    &\textbf{85.39}/0.05     &13.93/0.51/0.03   &0.01/0.04/0.04/$\sim 0$\\
                &$[1.11, -12.32, 0.99]$   &\ldots    &\textbf{73.36}/0.99     &25.60/0.70/0.09   &0.04/0.06/0.04/$\sim 0$\\\hline
$3/2(3/2^-)$    &$[2.80, -0.43, 4.27]$    &\ldots    &\textbf{96.40}/0.02     &2.07/0.06/0.40    &0.87/0.03/0.15/$\sim 0$\\
                &$[2.96, -3.66, 1.63]$    &\ldots    &\textbf{83.30}/0.10     &8.55/0.08/1.31    &5.87/0.13/0.630.02\\
                &$[3.12, -14.11, 0.77]$   &\ldots    &\textbf{53.33}/0.27     &19.43/0.02/2.40   &22.49/0.33/1.64/0.07\\\midrule[1pt]
&&&&$\Sigma_cD^*|{}^4S_{3/2}/{}^2D_{3/2}/{}^4D_{3/2}\rangle$        &$\Sigma_c^*D^*|{}^4S_{3/2}/{}^2D_{3/2}/{}^4D_{3/2}/{}^6D_{3/2}\rangle$\\
$1/2(3/2^-)$    &$[0.91, -0.65, 3.83]$   &\ldots &\ldots  &\textbf{97.47}/0.54/1.81    &0.16/$\sim 0$/0.01/$\sim 0$\\
                &$[0.96, -4.48, 1.72]$   &\ldots &\ldots  &\textbf{95.68}/0.84/2.93    &0.50/0.02/0.02/$\sim 0$\\
                &$[1.01, -12.10, 1.15]$  &\ldots &\ldots  &\textbf{94.74}/0.93/3.32    &0.95/0.03/0.02/$\sim 0$\\\hline
$3/2(3/2^-)$    &$[2.60, -1.50, 2.47]$   &\ldots &\ldots  &\textbf{82.12}/0.30/1.50    &15.35/0.09/0.59/0.03\\
                &$[2.70, -6.22, 1.16]$   &\ldots &\ldots  &\textbf{60.54}/0.38/1.89    &\textbf{35.59}/0.24/1.28/0.08\\
                &$[2.80, -14.48, 0.74]$  &\ldots &\ldots  &\textbf{44.06}/0.36/1.81    &\textbf{51.47}/0.38/1.80/0.13\\
\bottomrule[2pt]
\end{tabular}
\end{table*}

\subsection{The $\Sigma_c D$ system}

The isospin $I$ and spin $J$ for the $S-$wave $\Sigma_cD$ system
with negative parity are $(I, J) = (1/2, 1/2)$, $(3/2, 1/2)$. For $
J = 1/2$, the coupled channel wave function can be expanded as
\begin{eqnarray}
    |\Psi \rangle &=& \psi_1(r) \Sigma_cD|^2S_{1/2} \rangle + \psi_2(r) \Sigma_c^*D|^4D_{1/2} \rangle  \nonumber \\
    & + & \psi_3(r)\Sigma_cD^*|^2S_{1/2} \rangle  + \psi_4(r)\Sigma_cD^*|^4D_{1/2} \rangle \nonumber \\
    & + & \psi_5(r)\Sigma_c^*D^*|^2S_{1/2} \rangle + \psi_6(r) \Sigma_c^*D^*|^4D_{1/2} \rangle \nonumber \\
    & + & \psi_7(r) \Sigma_c^*D^*|^6D_{1/2} \rangle,
 \end{eqnarray}
with isospin $I = 1/2$ and $3/2$.

For the single channel $\Sigma_cD|^2S_{1/2} \rangle $ with isospin
$I = 1/2$, our results indicate that only the pion-exchange
potential is not strong enough to bind the $\Sigma_cD$ system as the
$DD\pi$ coupling is forbidden by the spin-parity conservation rule.
After taking into account the heavier $\rho$, $\omega$, and $\sigma$
exchanges accounting for the intermediate- and short-range
interactions, we obtain a loosely bound state with binding energy
$-0.23$ MeV and root-mean-square (rms) radius $5.21$~fm for a
reasonable cutoff $1.14$ MeV. As the cutoff increases to $1.34$~GeV,
the binding energy increases to $-18.18$~MeV while the rms radius
decreases to $0.97$ fm. Here, the intermediate- and short-range
forces from the OBE model play an important role to form the single
$\Sigma_cD$ molecular state with $I(J^P)=1/2(1/2^-)$.

When we include the coupled channel effects from channels
$\Sigma_cD^*$ and $\Sigma_c^*D^*$, we obtain a weakly bound  state
with binding energy $-0.87$~MeV and rms radius $3.41$ fm for cutoff
$\Lambda = 1.21$~GeV using only the long-range pion-exchange
potential. The probability of the dominant channel
$\Sigma_cD|^2S_{1/2} \rangle$ is $96.90\%$. Compared to the single
channel case, the coupled channel effects is helpful to form this
coupled doubly charmed molecular pentaquarks. If we further include
the intermediate- and short-range interaction from the heavier
$\rho$, $\omega$, $\sigma$, and $\eta$ exchanges, the cutoff
decreases to $0.96$ GeV to obtain a loosely bound state which has a
binding energy $-0.49$~MeV and rms radius $4.29$~fm which is twice
size of the deuteron. The dominant channel is $\Sigma_cD|^2S_{1/2}
\rangle$ with probability of $98.62\%$ and that for other channels
are less than $2\%$. As the cutoff increases to $1.06$ GeV, the
binding energy increases to $-15.47$ MeV while the rms radius
decreases to $1.03$ fm, which is still the size of the loosely bound
hadronic molecule. Here, our results indicate again the $\rho$,
$\omega$, $\sigma$, and $\eta$ exchanges do provide attractive
interactions in binding the $\Sigma_cD$ bound state with
$1/2(1/2^-)$. To summarize, we propose the $\Sigma_cD$ to be a good
candidate of hadronic molecular state.

For the isospin $I = 3/2$ case, we could obtain bound state for the
single channel $\Sigma_cD$ neither with the OPE potential nor with
the OBE potential. After we include the coupled channel effects from
channels $\Sigma_c^* D$ as well as $\Sigma_cD^*$, a loosely bound
state with binding energy $-0.64$~MeV and rms radius $3.65$~fm
appears when the cutoff is tuned to $1.62$~GeV using the OPE
potential. We further include the intermediate- and short-range
interaction from the heavier $\rho$, $\omega$, $\sigma$, and $\eta$
exchanges, a loosely bound state is obtained with binding energy
$-0.53$~MeV and rms radius $4.05$~fm. The probability of
$\Sigma_cD|^2S_{1/2} \rangle$ is $97.78\%$ and that of
$\Sigma_c^*D^*|^2S_{1/2} \rangle$ is $1.32\%$. The probabilities for
other channels are very tiny, less than $1\%$. As the cutoff
increases to $1.95$ GeV, the binding energy increases to
$-13.52$~MeV, and the rms radius decreases to $0.92$ fm. Meanwhile,
more coupled channel effects get involved. The probability of
$\Sigma_cD|^2S_{1/2} \rangle$ is $82.43\%$ while that for
$\Sigma_c^*D^*|^2S_{1/2} \rangle$ is $13.62\%$. The probabilities
for other channels are small, less than $3\%$. From the current
numerical results, the system $\Sigma_cD[I(J^P) = 3/2(1/2^-)]$ may
also be viewed as a candidate of doubly charmed hadronic molecule.

\subsection{The $\Sigma_C^* D$ system}

Due to the higher threshold of $\Sigma_c^*D$ compared to
$\Sigma_cD$, we perform a calculation relative to the threshold of
$\Sigma_c^*D$. The isospin $I$ and spin $J$ for $\Sigma_c^*D$ with
negative parity can be $(I, J) = (1/2, 3/2)$ and  $(3/2, 3/2)$. The
coupled channel wave function can be expanded as
\begin{eqnarray}
    |\Psi \rangle &=& \psi_1(r) \Sigma_c^*D|^4S_{3/2} \rangle
    + \psi_2(r) \Sigma_c^*D|^4D_{3/2} \rangle  \nonumber \\
    & + & \psi_3(r)\Sigma_cD^*|^4S_{3/2} \rangle
    + \psi_4(r)\Sigma_cD^*|^2D_{3/2} \rangle \nonumber \\
    & + & \psi_5(r)\Sigma_cD^*|^4D_{3/2} \rangle
    +  \psi_6(r)\Sigma_c^*D^*|^4S_{3/2} \rangle  \nonumber \\
    & + & \psi_7(r) \Sigma_c^*D^*|^2D_{3/2} \rangle
     +  \psi_8(r) \Sigma_c^*D^*|^4D_{3/2} \rangle \nonumber \\
    & + & \psi_9(r) \Sigma_c^*D^*|^6D_{3/2} \rangle,
 \end{eqnarray}
with isospin $I = 1/2$ and $3/2$.

In the heavy quark limit, the OBE effective potentials for the
$S-$wave $\Sigma_c^*{D}$ system are very similar to the $S-$wave
$\Sigma_cD$ interactions. In the isospin $I = 1/2$ case, we could
not obtain bound state solutions for the single $\Sigma_c^*D$ with
the long-range pion exchange potential only. After we take into
account the intermediate- and short-range interaction from the
heavier $\rho$, $\omega$, $\sigma$, and $\eta$ exchanges, a loosely
bound state with binding energy $-0.57$~MeV and rms radius $4.05$~fm
emerges. If we include the coupled channel effects from
$\Sigma_cD^*$ and $\Sigma_c^*D^*$ channels, we obtain a loosely
bound state with the binding energy $-0.76$~MeV and the rms radius
$3.09$~fm using the only long-range pion exchange when the cutoff is
set to $1.48$~GeV. When we further include the intermediate- and
short-range interactions from the heavier $\rho$, $\omega$,
$\sigma$, and $\eta$ exchanges, we obtain a loosely bound state with
a smaller cutoff $1.05$~GeV which is comparable to the value used
for the study of deuteron \cite{Tornqvist:1993ng,Tornqvist:1993vu}.
Its binding energy is $-1.30$~MeV and rms radius is $2.84$~fm. The
probability for the dominant channel $\Sigma_c^*D|^4S_{3/2} \rangle$
is $94.50\%$ and that for $\Sigma_cD^*|^4S_{3/2}\rangle$ is
$5.14\%$. As the cutoff is tuned to $1.11$~GeV, the binding energy
becomes $-12.32$~MeV and rms radius decreases to $0.99$~fm. The
probability for the $\Sigma_c^*D|^4S_{3/2}\rangle$ channel decreases
to $73.36\%$ whereas that for the $\Sigma_cD^*|^4S_{3/2}\rangle$
channel increases to $25.60\%$ due to the very close threshold for
these two channels. In fact, there exists competition between the
binding energy and the threshold difference for the channels
involved. When the binding energy is small, the role of the
threshold difference between the channels involved will be
amplified. However, when the binding energy becomes bigger, this
effect will disappear. For a conclusion, the $\Sigma_c^*D$ is a good
candidate of a hadronic molecule in the present OBE potential model.

For the isospin $I = 3/2$ case, we only obtain bound state solutions
with the cutoff larger than $2.0$~GeV which is larger than the
normal value used for the study of the deuteron and other hadronic
molecule within the OBE model. Thus, it is less likely that the
$\Sigma_c^*D$ with $3/2(3/2^-)$ can form a loosely bound hadronic
molecule.

\subsection{The $\Sigma_c D^*$ system}

The isospin $I$ and spin $J$ for the system $\Sigma_cD^*$ with
negative parity can be $(I, J) = (1/2, 1/2)$, $(1/2, 3/2)$, $(3/2,
1/2)$, and $(3/2, 3/2)$. For the spin $J = 1/2$ case, the coupled
channel wave function can be expanded as
\begin{eqnarray}
    |\Psi \rangle &=& \psi_1(r) \Sigma_cD^*|^2S_{1/2} \rangle
    + \psi_2(r) \Sigma_cD^*|^4D_{1/2} \rangle  \nonumber \\
    & + & \psi_3(r)\Sigma_c^*D^*|^2S_{1/2} \rangle
    +  \psi_4(r)\Sigma_c^*D^*|^4D_{1/2} \rangle  \nonumber \\
    & + & \psi_5(r) \Sigma_c^*D^*|^6D_{1/2} \rangle
 \end{eqnarray}
for isospin $I = 1/2$ and $3/2$.

For the $(I, J) = (1/2, 1/2)$ case, we find bound states solutions
for the single channel $\Sigma_cD^*$ using the long-range pion
exchange potential only when the cutoff is tuned to be larger than
$2.40$ GeV. After taking into account the intermediate- and
short-range $\rho$, $\omega$, $\sigma$, and $\eta$ exchanges, a
loosely bound state is obtained with smaller cutoff $1.30$ GeV. Its
binding energy is $-0.51$~MeV, and the rms radius is $4.41$ fm. If
we include the coupled channel effects instead of the intermediate-
and short-range interactions, we obtain bound state solutions with
cutoff tuned larger than $2.24$ GeV. However, after we take into
account the coupled channel effects as well the heavier $\rho$,
$\omega$, $\sigma$, and $\eta$ exchanges, a loosely bound state
emerges with a smaller cutoff $1.27$~GeV, which is comparable to the
value used for the study of deuteron and other hadronic molecules
within the OBE model. Its binding energy is $-0.41$~MeV and the rms
radius $4.70$~MeV. The probability of $\Sigma_cD^*|^2S_{1/2}
\rangle$ is $97.56\%$ while that of $\Sigma_cD^*|^4D_{1/2} \rangle$
is $2.24\%$, and that for other channels is tiny. As the cutoff
increases to $1.47$ GeV, the binding energy increases to
$-11.22$~MeV and rms radius decreases to $1.30$~fm. Meanwhile, the
probability of the S-wave channel $\Sigma_cD^*|^2S_{1/2} \rangle$
decreases to $88.46\%$ and that of the D-wave channel
$\Sigma_cD^*|^4D_{1/2} \rangle$ increases to $9.44\%$. From the
current numerical results, the system $\Sigma_cD^*$ with $(I, J) =
(1/2, 1/2)$ should be viewed as a good candidate of double charm
molecular pentaquark, and the intermediate- and short-range
interactions from the $\rho$, $\omega$, $\sigma$, and $\eta$
exchanges are very important, especially in the single channel case.

For the $(I, J) = (3/2, 1/2)$ case, there exist bound state
solutions with either OPE potential or OBE potential. When we take
into account the coupled channel effects as well as the
intermediate- and short-range interaction from the heavier $\rho$,
$\omega$, $\sigma$ and $\eta$ exchanges, a loosely bound state
appears for the cutoff tuned to $1.41$~GeV. Its binding energy is
$-0.48$~MeV and the rms radius is $4.09$~fm. It is almost a pure
$\Sigma_cD^*|^2S_{1/2} \rangle$ bound state with its probability of
$99.53\%$. As the cutoff increases to $1.81$~GeV, the binding energy
increases to $-17.34$~MeV and the rms radius decreases to $0.84$~fm.
The probability of $\Sigma_cD|^2S_{1/2}\rangle$ decreases to
$91.13\%$ while that of the $\Sigma_c^*D^*|^2S_{1/2} \rangle$
channel increases to $8.04\%$. The reasonable cutoff parameter and
the bound state properties support the $\Sigma_cD^*$ state with
$3/2(1/2^-)$ as a candidate of the doubly charmed molecular
pentaquark.

For the spin $J = 3/2$ case, the coupled channel wave function can
be expanded as
\begin{eqnarray}
    |\Psi \rangle &=& \psi_1(r) \Sigma_cD^*|^4S_{3/2} \rangle
    + \psi_2(r) \Sigma_cD^*|^2D_{3/2} \rangle  \nonumber \\
   & + & \psi_3(r) \Sigma_cD^*|^4D_{3/2} \rangle
    +  \psi_4(r)\Sigma_c^*D^*|^4S_{1/2} \rangle  \nonumber \\
   & + & \psi_5(r) \Sigma_c^*D^*|^2D_{1/2} \rangle
    +  \psi_6(r)\Sigma_c^*D^*|^4D_{3/2} \rangle  \nonumber \\
    & + & \psi_7(r) \Sigma_c^*D^*|^6D_{3/2} \rangle
 \end{eqnarray}
for isospin $I = 1/2$ and $3/2$.

The system $\Sigma_cD^*$ with $(I, J) = (1/2, 3/2)$ is very
interesting. We obtain a loosely bound state with only the
long-range pion-exchange potential for the cutoff $1.20$~GeV and
with the OBE potential for the cutoff $0.91$~GeV. When we take into
account the coupled channel effects as well as the intermediate- and
short-range interaction from the heavier $\rho$, $\omega$, $\sigma$,
and $\eta$ exchanges, a loosely bound state is obtained with cutoff
$0.91$~GeV. The probability for the dominant channel
$\Sigma_cD^*|^4S_{3/2} \rangle$ is $97.47\%$ while that of
$\Sigma_cD^*|^4D_{3/2} \rangle$ is $1.81\%$. As the cutoff increase
to $1.01$~GeV, the binding energy increases to $-12.10$~MeV while
the rms radius decreases to $1.15$~fm which is still comparable to
the size of well-known deuteron. The probability of
$\Sigma_cD^*|^4S_{3/2} \rangle$ is $94.74\%$ and that for
$\Sigma_cD^*|^4D_{3/2} \rangle$ is $3.32\%$. From the current
numerical results, we propose the $\Sigma_cD^*$ state with
$1/2(3/2^-)$ as a good candidate of doubly charmed molecular
pentaquark.

For the isospin $I = 3/2$ case, we could not obtain bound state
solutions for the single channel $\Sigma_cD^*|^4S_{3/2} \rangle$
until tuning the cutoff to be as large as $4.0$~GeV for the
long-range pion-exchange potential only and $3.90$~GeV for the OBE
potential. Although, after both the coupled channel effects and the
intermediate- and short-range interactions are included, the cutoff
with the bound state solutions is still too larger compared the
reasonable value, which is around $1.0$ GeV. Thus, the $\Sigma_cD^*$
system with $(I, J) = (3/2, 3/2)$ may not be a candidate of hadronic
molecule.

To draw a conclusion for the $\Sigma_cD^*$ systems with
$1/2(1/2^-)$, $1/2(3/2^-)$, $3/2(1/2^-)$, and $3/2(3/2^-)$, we
propose the $\Sigma_cD^*$ states with $1/2(1/2^-)$ and $1/2(3/2^-)$
to be good candidates of the doubly charmed molecular pentaquarks.
The $\Sigma_cD^*$ with $3/2(1/2^-)$ may also be viewed as a
candidate of hadronic molecule whereas the $\Sigma_cD^*$ with
$3/2(3/2^-)$ is not supported to be a candidate of a hadronic
molecule.

\section{Summary}\label{sec4}

We perform a systematic exploration of the possible doubly charmed
molecular pentaquark of $\Sigma_c^{(*)}D^{(*)}$ with the
one-boson-exchange potential model. To investigate the the $S-D$
wave mixing effects, the coupled channel effects, the long-rang pion
exchange interaction and the intermediate- and short-range
interactions arising from the heavier $\rho$, $\omega$, $\sigma$,
and $\eta$ exchanges, we performed four kinds of calculations. We
first do the calculation for the single channel using the OPE
potential and OBE potential individually. Then we include the
coupled channel effects and do the calculations again within the OPE
potential model. Finally, we take into account all the effects and
do a full coupled channel calculation within the OBE potential
model. We also find that for a loosely bound states in the coupled
channel study, the binding energy and the threshold difference
between different channels compete with each other. In other words,
for a loosely bound state, the role of the threshold difference will
be amplified by the small binding energy, which had already been
emphasized in the study of the isospin breaking for the $X(3872)$
\cite{Li:2012cs} and $T_{cc}$ \cite{Chen:2021vhg}.

Our results reveal some general features for the coupled channel
study of the $\Sigma_c^{(*)}D^{(*)}$ systems with the OBE model,
which has already been discovered in previous coupled channel study
of the hadronic molecules \cite{Chen:2017xat,Chen:2018pzd}. Overall,
the coupled channel effects are helpful for the formation of the
bound states. The long-range pion-exchange potential plays an
important role in the formation of the loosely bound states while
the intermediate- and short-range interaction from the heavier
$\rho$, $\omega$, $\sigma$, and $\eta$ exchanges can also help
strengthen the binding between $\Sigma_c^{(*)}$ and $D^{(*)}$. Very
interestingly, we propose some good candidates of the doubly charmed
molecular pentaquarks. From our results, the $\Sigma_cD$ state with
$I(J^P) = 1/2(1/2^-)$, $\Sigma_c^*D$ state with $I(J^P) =
1/2(3/2^-)$, and $\Sigma_cD^*$ states with $1/2(1/2^-)$,
$1/2(3/2^-)$ are good candidates of doubly charmed hadronic
molecules. The $\Sigma_cD$ state with $3/2(1/2^-)$ and $\Sigma_cD^*$
with $3/2(1/2^-)$ may also be viewed as the doubly charmed molecular
candidates. We also find that the $\Sigma_c^*D$ state with
$1/2(3/2^-)$ is more complicated due to the near threshold between
the $\Sigma_cD^*$ and $\Sigma_c^*D$ systems. For a loosely bound
state with the binding energy around $1$~MeV, the dominant channel
is the $\Sigma_c^*D|^4S_{3/2}\rangle$ with the probability larger
than $95\%$ and small contributions from the
$\Sigma_cD^*|^4S_{3/2}\rangle$ channel. When the binding energy is
around tens of MeV, the probability of the
$\Sigma_c^*D|^4S_{3/2}\rangle$ will be comparable with that of the
$\Sigma_cD^*|^4S_{3/2} \rangle$ channel.

The newly observed $T_{cc}$ with doubly charmed would definitely be
a new hadronic state beyond the traditional baryons and mesons. Its
observation opens a new window to search for the new hadronic state
experimentally, and indicates there comes to a new era for the
experimental research of the exotic hadronic states. The $\Sigma_cD$
and $\Sigma_c^*D$ states can be searched for by analyzing the
$\Lambda_cD\pi$ invariant mass spectrum of the bottom baryon and $B$
meson decays. The $\Sigma_cD^*$ states can be searched for in the
invariant mass spectrum of $\Lambda_cD^*\pi$, $\Lambda_cD\pi\pi$ and
$\Lambda_cD\pi\gamma$. Since the width of $\Sigma_c^*$ is much
larger than that of $D^*$, $\Sigma_c^*D\rightarrow \Lambda_cD\pi$
would be the dominant decay mode. We sincerely hope these proposed
doubly charmed molecular candidates will be searched for by the LHCb
or BelleII Collaborations in the near future.

\section*{ACKNOWLEDGMENTS}
This work is supported by the National Natural Science Foundation of
China under Grants 11975033 and 12070131001, the China National
Funds for Distinguished Young Scientists under Grant No. 11825503,
National Key Research and Development Program of China under
Contract No. 2020YFA0406400, and the 111 Project under Grant No.
B20063, the Fundamental Research Funds for the Central Universities
under Grants No. lzujbky-2021-sp24. R. C. is supported by the
National Postdoctoral Program for Innovative Talent.

\end{document}